# Anisotropic van der Waals Crystal with High Refractive Index and Transparency for UV-Visible Range Applications

*Aleksandr S. Slavich[1†], Georgy A. Ermolaev[1†], Ilya A. Zavidovskiy[2†], Dmitriy V. Grudinin[1†], Konstantin V. Kravtsov[1†], Mikhail K. Tatmyshevskiy[2], Mikhail S. Mironov[1], Adilet N. Toksumakov[1], Gleb I. Tselikov[1], Ilia M. Fradkin[1], Kirill V. Voronin[3], Maksim R. Povolotskiy[2], Olga G. Matveeva[2], Alexander V. Syuy[1], Dmitry I. Yakubovsky[2], Dmitry M. Tsymbarenko[4], Ivan Kruglov[1], Davit A. Ghazaryan[2,5], Sergey M. Novikov[2], Andrey A. Vyshnevyy[1], Aleksey V. Arsenin[1,5], Valentyn S. Volkov[1,5\*], and Kostya S. Novoselov[6,7,8\*]*

[1]*Emerging Technologies Research Center, XPANCEO, Internet City, Emmay Tower, Dubai, United Arab Emirates*

[2]*Moscow Center for Advanced Studies, Kulakova str. 20, Moscow, 123592, Russia*

[3]*Donostia International Physics Center (DIPC), Donostia/San Sebastián, 20018, Spain*

[4]*Department of Chemistry, Lomonosov Moscow State University, Moscow, 119991, Russia*

[5]*Laboratory of Advanced Functional Materials, Yerevan State University, Yerevan 0025, Armenia*

[6]*National Graphene Institute (NGI), University of Manchester, Manchester, M13 9PL, UK*

[7]*Department of Materials Science and Engineering, National University of Singapore, Singapore, 03-09 EA, Singapore*

[8]*Institute for Functional Intelligent Materials, National University of Singapore, 117544, Singapore, Singapore*

[†]These authors contributed equally to this work

\*Correspondence should be addressed to e-mail: kostya@nus.edu.sg and vsv@xpanceo.com

**Abstract**

**Thanks to their record high refractive index and giant optical anisotropy, van der Waals (vdW) materials have accelerated the development of nanophotonics. However, traditional high refractive index materials, such as titanium dioxide (TiO$_2$), still dominate in the most important visible range. This is due to the current lack of transparent vdW materials across the entire visible spectrum. In this context, we propose that germanium disulfide (GeS$_2$) could offer a significant breakthrough. With its high refractive index, negligible losses, and biaxial optical anisotropy across the whole visible range, GeS$_2$ has the potential to complement TiO$_2$ and close the application gap of vdW materials in the visible spectrum. The addition of GeS$_2$ could have a profound impact on the design of van der Waals nanophotonic circuits for any operation wavelength from ultraviolet to infrared, emphasizing the significance of the potential impact of GeS$_2$ on the field of nanophotonics.**

**Introduction**

Van der Waals (vdW) crystals have received considerable interest since their first appearance[1]. Their captivating optical properties[2–5] provided the groundwork for modern photonic applications, such as extreme skin-depth guiding[2,6], topological phase singularities[7,8], Mie nanoresonators[9,10], exciton-polaritons[11,12], to name a few. Moreover, recent studies[13–15] demonstrated that low-symmetry vdW crystals, such as As$_2$S$_3$[13], α-MoO$_3$[14], and black phosphorus[15], exhibit strong in-plane optical and mechanical anisotropy, which enables novel photonic effects and devices, including ultrathin waveplates[13,16], twisted nanooptics[17–19], polarization-sensitive photodetectors[20,21], and wandering principal optical axes[22,23]. The core parameter in all these applications is the refractive index and its unique dependence on wavelength and direction, where all records of absolute values belong to vdW materials[3]. As a result, several groups[3,24,25] view vdW crystals as a perfect material platform for the next-generation nanophotonics. Hoverwer, a smooth transition from conventional high refractive index materials, such as Si[26], GaP[27], and TiO$_2$[28], to vdW materials is hindered by the absence of high refractive index vdW crystal transparent in the entire visible spectrum.

One promising crystals is germanium disulfide (GeS$_2$), which stands out with low-symmetry crystallography[29] and a large bandgap of 3.4 – 3.7 eV, which lies at the boundary between visible and ultraviolet (UV) spectral intervals[30–32]. Amorphous GeS$_2$ has already proven its utility in versatile photonic applications, such as chalcogenides glasses in optical fibers[33], photodoping[34], and gradient refractive index Fresnel lenses[35]. With a plethora of distinguished properties, layered GeS$_2$ also emerged as a focal point for fundamental and applied research[36–39]. However, the anisotropic optical properties of GeS$_2$ have not yet been investigated thoroughly, making it a crucial area for further exploration.

This work introduces GeS$_2$ as a new wide-bandgap vdW material with a large in-plane anisotropy. We present a comprehensive examination of its anisotropic structural, vibrational, and optical properties. First, we performed a detailed study of GeS$_2$ crystal structure and vibrational modes, establishing the relationship between those. Then we probed the full dielectric tensor of GeS$_2$ from ultraviolet (UV) to near-infrared (NIR) range by spectroscopic ellipsometry and underpinned our findings by first-principle calculations of dielectric function and vibrational modes. Finally, we independently verified the acquired optical constants through the characterization of waveguide modes propagating in GeS$_2$ using s-SNOM technique. Our work shows that GeS$_2$ demonstrates high optical anisotropy and the highest refractive index among transparent vdW materials in the visible-UV range down to 360 nm, making it, thus far, the only highly refractive anisotropic vdW material without optical losses across the entire visible spectrum.

## Results

**Anisotropic crystal structure of van der Waals GeS$_2$**

Germanium disulfide is a gray semiconducting layered crystal from group IV-VI metal dichalcogenides. Figure 1a shows a schematic illustration of the monoclinic GeS$_2$ crystal structure. GeS$_2$ unit cell consists of two layers weakly bonded by van der Waals forces[29]. At the same time, each layer is formed by combining visually distinctive structural units. The first block is a chain of tetrahedra GeS$_4$ connected by a common corner. Shared sulfur atoms are referred to as "corner S-atoms"[40] (marked by red rectangle). The second block is formed by two edge-shared tetrahedral units of GeS$_4$ (marked by a blue quadrilateral). These units form Ge-S-Ge-S rings whose S atoms we will call the "edge" S-atoms. The S-atoms connecting these two blocks are called the "bridges" S-atoms and are marked by green circles. Considering all these three patterns together (see colored dashed lines in Figure 1a), it becomes apparent that the neighboring adjacent set (inside a given sublayer) is a copy of the initial set inverted relative to the *b*-axis. This means that the *b*-axis is a screw axis of $2_1/c$. As shown in Figure 1a, the resulting complex structure of the individual monolayer is riddled with infinite-shape hollows. These facts make GeS$_2$ one of the most interesting materials from the crystallographic point of view.

Figure 1b represents an optical image of GeS$_2$ bulk crystal. Since precise crystallographic information is critically important for *ab initio* calculations, we commenced our study by applying the X-ray diffraction (XRD) technique to refine its lattice constants. Our XRD results indicate $P2_1/c$ space group with lattice parameters of $a = 6.6946(14)$ Å, $b = 16.037(3)$ Å, $c = 11.423(2)$ Å, $\alpha = 90°$, $\beta = 90°$, $\gamma = 90°$ giving a remarkable accuracy with *R*-value of 3.6%. For further details on crystallographic data collection, see Methods and Supplementary Note 1. Additionally, we verified the crystal structure of thin GeS$_2$ by selected area electron diffraction (see Supplementary Figure 3a-c). To ensure the quality of the flake, we investigated its elemental composition by energy-dispersive X-ray spectroscopy (EDS). From EDS, we found atomic stoichiometry of Ge:S ≈ 35.9/64.1 %, which within the measurement accuracy corresponds to GeS$_2$.

The bulk crystal is amenable to exfoliation into a single layer, as shown in Figure 1c. The thickness of the monolayer is half of the unit cell along the *c*-axis (t ≈ 0.5 × 1.14 nm = 0.57 nm). Considering that the height step observed *via* AFM is less than twice this thickness, we can confirm the exfoliation of a monolayer of GeS$_2$ (see inset in Figure 1c). A similar discrepancy between measured (0.87 nm) and expected thicknesses in AFM studies of other 2D materials has been attributed to tip-surface interactions and surface chemistry[41].

Based on these crystallographic data, we estimate dielectric properties and the bandstructure of the material *via* first-principles calculations (see Methods and Supplementary Note 2). Remarkably, GeS$_2$ possesses one of the largest bandgaps within the layered semiconductor family while maintaining a relatively high out-of-plane *static* dielectric permittivity (see Figure 1d). Hence, GeS$_2$ can serve as an alternative insulating material to hBN in van der Waals nanoelectronics[42]. The GeS$_2$ bandstructure cuts along crystallographic directions determined by *ab initio* calculations are presented in Figure 1e-

g. The differences in dispersions for different directions as well as the flat bands along the *c*-axis indicate a strong anisotropic dielectric response, making GeS$_2$ a perfect candidate for optical studies.

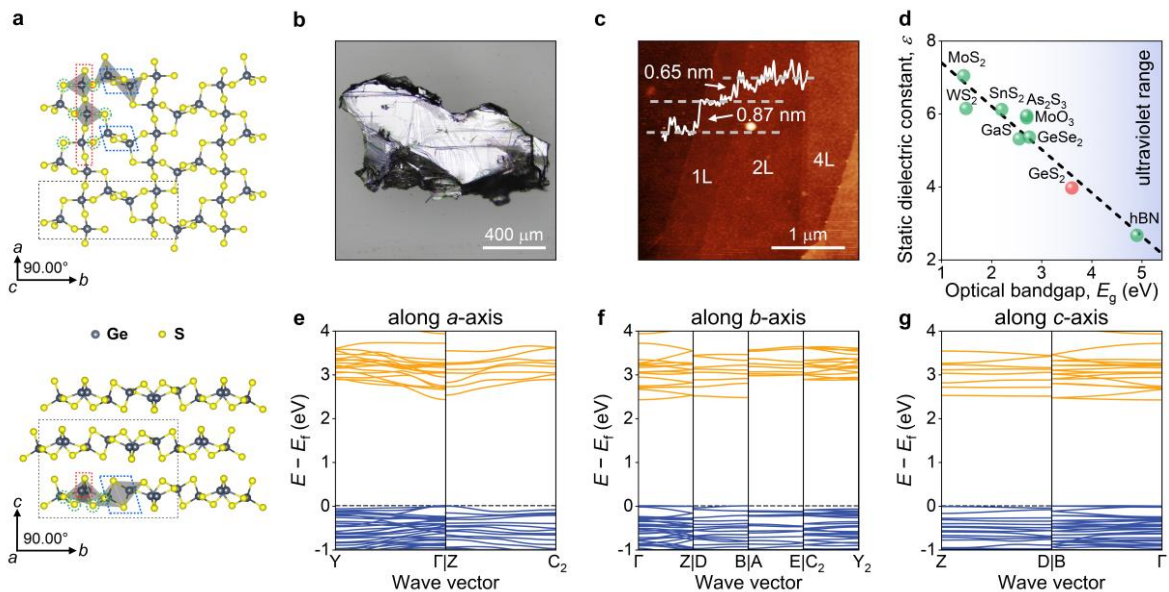

**Figure 1. Crystal structure of van der Waals GeS$_2$ and fingerprints of its anisotropy. (a)** The monoclinic crystal structure of the GeS$_2$ monolayer along the *c*-axis (top), and the bulk structure along the *a*-axis (bottom). A black dashed frame reveals the unit cell. Colored dashed lines encircle selected groups of atoms that define vibrational interaction: red – corner-sharing patterns, blue – edge-sharing patterns, green – bridge sulfur atoms **(b)** Optical image of GeS$_2$ bulk crystal. **(c)** AFM scan of GeS$_2$ flake, revealing the presence of GeS$_2$ monolayer. **(d)** GeS$_2$ in the framework of the calculated out-of-plane static dielectric constants of novel van der Waals materials *vs* their experimental optical bandgaps. The electronic bandstructure cuts of GeS$_2$ along *a*-axis **(e)**, *b*-axis **(f)**, and *c*-axis **(g)**. Orange and blue curves present conduction and valence bands, respectively.

**GeS$_2$ anisotropic vibrational response**

Having refined the crystallographic properties of GeS$_2$, we proceed with the determination of its anisotropic vibrational response. In doing so, we pursue several goals. First, polarized Raman spectra are one of the material fingerprints, that contain information regarding defects, crystallographic structure, and orientation in numerous vdW materials[43–52]. Although previous works have demonstrated Raman response and theoretical analysis of vibrational modes in vdW GeS$_2$[32,40,53,54], the origin of the most observed Raman lines in GeS$_2$ and their anisotropic nature remains open for further investigation. Next, *ab initio* calculations of vibrational response provide additional verification for other theoretical calculations, in particular, of optical properties. Considering these factors, the detailed experimental and theoretical study of anisotropic optical and vibrational properties of GeS$_2$ is of great interest.

To explore the vibrational signature of GeS$_2$ and identify the orientation of the crystallographic axes of the sample, we utilized the polarization-resolved Raman spectroscopy method in combination with theoretical analysis of the vibrational modes. Here, we acquired Raman spectra for different polarization angles from 0° to 360° with the step of 10° by rolling the sample between a parallel-oriented polarizer and analyzer. Then, we used a phonon-based approach to calculate the vibrational response of the whole atomic system[55] (see Supplementary Note 2). A detailed discussion of vibration analysis is provided in Supplementary Note 3. Figure 2a presents the Raman spectra derived by averaging over all angles for 532 nm and 632.8 nm excitation wavelengths, as well as DFT vibrational mode calculations utilizing crystallographic parameters built upon our XRD measurements.

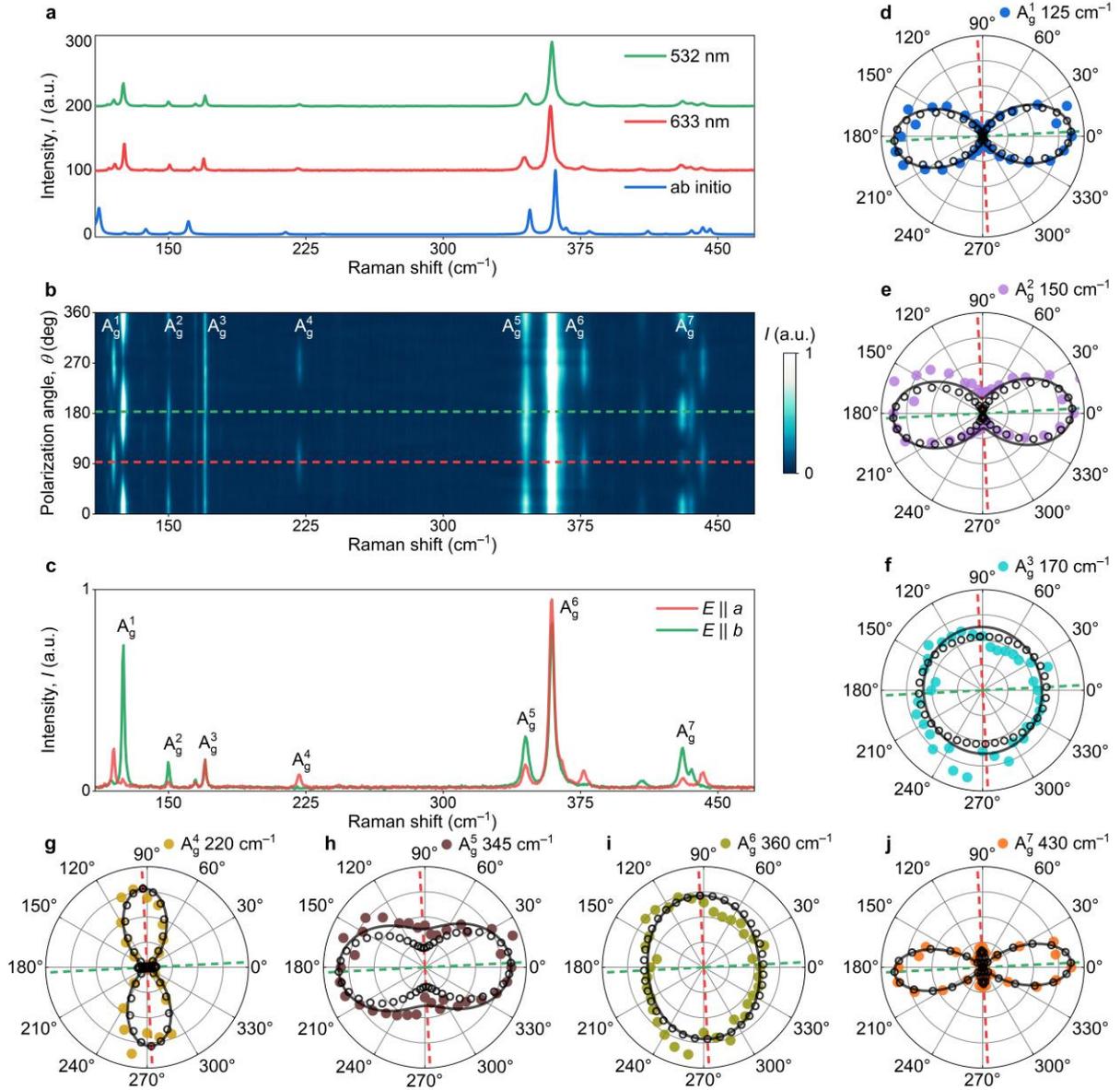

**Figure 2. Unveiling the anisotropic properties of van der Waals GeS$_2$ with polarized Raman spectroscopy.** (**a**) Raman spectra of GeS$_2$ measured at excitation wavelengths of 532 nm and 632.8 nm along with the calculated spectra. (**b**) Angle-resolved polarized Raman intensity colormap of bulk GeS$_2$ (thickness, $t$ = 151 nm) acquired by rotating the sample in a parallel-polarized configuration. (**c**) Raman spectra along the two crystalline axes (*a* and *b*) of GeS$_2$. (**d-j**) Polar plots of the fitted peak intensities for 7 Raman modes. Experimental data is shown by filled circles, its fitting is indicated by solid lines. DFT data is presented by hollow circles. Dashed colored lines indicate the direction of the crystallographic axes of the studied sample.

Figure 2b displays the angle-dependent polarized Raman intensity colormap for 151 nm-thick GeS$_2$ flake on Schott glass substrate. The calculated angle-averaged spectrum demonstrates a high alignment with the experimentally acquired spectrum, as shown in Figure 2a. This result is further confirmed by adequate alignment between experimentally derived and *ab initio* angle-resolved intensities of 22 Raman-active lines, as indicated by polar plots presented in Supplementary Figure 10 (for 532 nm-excited spectra) and Supplementary Figure 11 (for 632.8 nm-excited spectra). As suggested in[40], bands in the low-wavenumber region originate from the Ge-S bending, while the high-wavenumber region is mainly attributed to the Ge-S bond stretching. Our vibrational mode analysis significantly extends this approach by elaborating on the origin of 13 lines. Among these 13 lines, the origin of 10 lines is reported for the first time. Close examination of Figure 2b reveals that modes in the range 90–180 cm$^{-1}$ are indeed assigned to various bending motions of Ge-S bonds, the 220-cm$^{-1}$-centered band is attributed to the breathing mode of (GeS)$_2$ rings, while 340–450 cm$^{-1}$ peaks are mainly ascribed to the stretching oscillations of S atoms. In such high-wavenumber oscillations, Ge atoms remain relatively immobile due to their high atomic mass in comparison to S atoms (see mode visualization in Supplementary Table 5,6).

Excellent corroboration between experimental and theoretical studies confirms that the suggested approach can be applied to predict Raman spectra and angle-resolved vibrational response of vdW materials. The calculated Raman spectra for incident laser polarization along two crystallographic axes, and the intensities of Raman-active modes as a function of polarization angle are shown in Supplementary Figure 8. The prominent in-plane anisotropy of germanium disulfide yields notable angular intensity variation for most lines. Acknowledging the fixed patterns of polar diagrams in relation to in-plane crystallographic directions, we analyzed polar plots for five intensive modes to determine the orientation of the crystallographic axes. We fitted the experimental data and aligned the obtained curves with the calculated polar plots. The fitting model is discussed in detail in Supplementary Note 3. The resulting Raman spectra for incident laser polarization along the two crystallographic axes are shown in Figure 2c. Figure 2d-k demonstrates an outstanding match between the experimental and calculated polar diagrams for $A_g^1 - A_g^7$ modes. When the incident laser polarization is aligned with the *b*-axis of GeS$_2$, Raman peaks at 125 ($A_g^1$), 150 ($A_g^2$), 345 ($A_g^5$), and 430 ($A_g^7$) cm$^{-1}$ have maximum intensity. Similarly, the *a*-axis corresponds to the maximum of 220 ($A_g^4$) and 360 ($A_g^6$) cm$^{-1}$ (although the low aspect ratio of 360 cm$^{-1}$ polar plot shape hinders its applicability in the determination of crystallographic axes). Our analysis reveals that $A_g^6$ vibration is synphase breathing of all GeS$_4$ tetrahedra. This explains why the mode has the highest intensity: light is scattered by simultaneously breathing tetrahedra. Hence, its Raman response is relatively isotropic, which is uncommon for anisotropic material (see Supplementary Table 5). Owing to the high intensity of $A_g^1 - A_g^7$ peaks, prominent anisotropy of their angle-resolved intensities, and their perfect match with calculated data, we propose these peaks for potential use in the determination of the crystallographic axes for layered GeS$_2$.

**Optical anisotropy of van der Waals GeS$_2$**

Generally, for monoclinic crystals, the dielectric tensor has a very nontrivial form [22,23,56]. Indeed, since the crystallographic axes are non-orthogonal - the dielectric tensor in Cartesian representation contains mixed contributions from different crystallographic axes, making extraction of underlying optical constants challenging. Fortunately, the monoclinic angle *β* of GeS$_2$, with a value of 90.935(7)°, is practically indistinguishable from 90°, allowing us to treat GeS$_2$ as an orthorhombic crystal. Hence, we can use the diagonal form of the GeS$_2$ dielectric tensor, neglecting its off-diagonal elements.

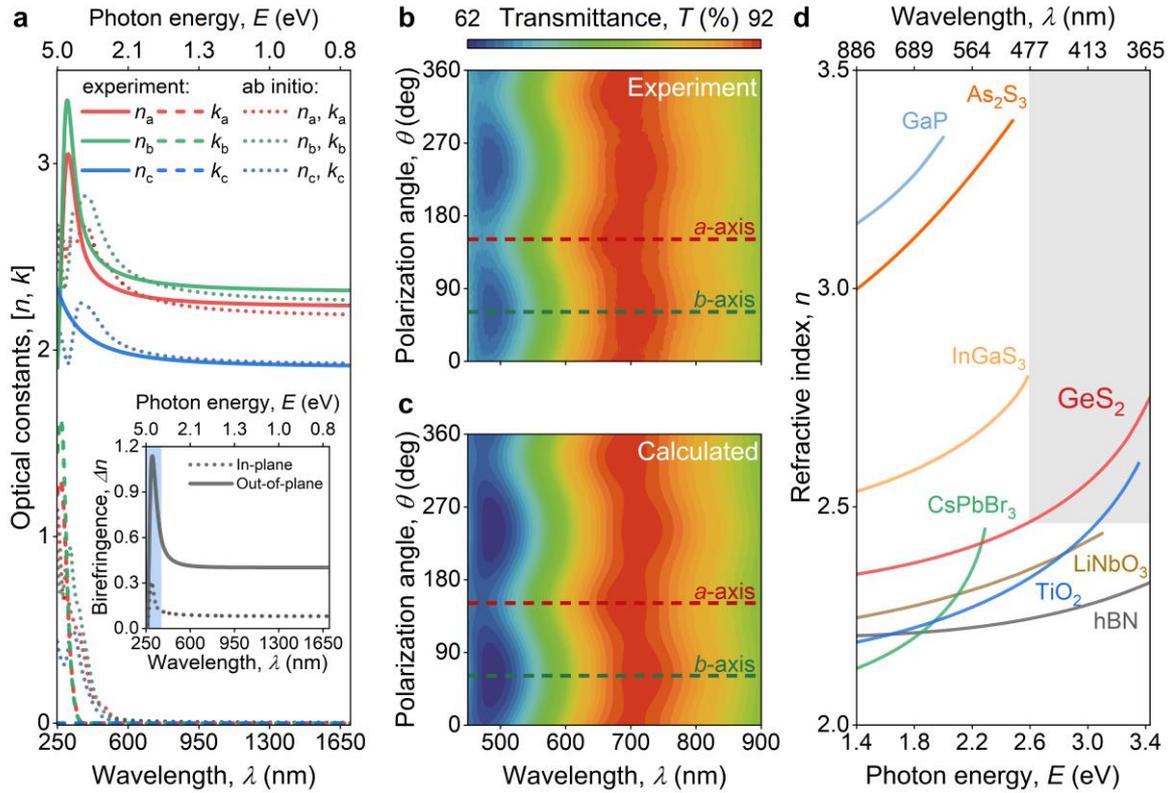

**Figure 3. Optically anisotropic van der Waals GeS$_2$ in the family of high refractive index materials.** (**a**) Optical constants of GeS$_2$ crystal, obtained *via* spectral ellipsometry. The inset shows the resulting in-plane (dotted line) and out-of-plane (solid line) birefringence. The shaded blue region indicates the spectral range with optical losses. (**b,c**) Experimental and calculated polarized micro-transmittance of GeS$_2$ obtained in parallel-polarized configuration. The dashed lines show the position of *a* (red) and *b* (green) crystallographic axes. (**d**) The comparison of the refractive indices of van der Waals GeS$_2$ with traditional photonic materials in their respective transparency windows.

We exfoliated GeS$_2$ flakes onto Si and Schott glass substrates, initially determining GeS$_2$ crystallographic axes by angle-resolved polarized Raman spectroscopy. Then, we performed spectroscopic ellipsometry measurements (see Methods and Supplementary Note 4 for more details). For enhanced accuracy, we analyzed data from multiple flakes with different thicknesses. The optical constants acquired from the ellipsometry data spectra are depicted in Figure 3a. The inset in Figure 3a presents the birefringence values between the in-plane refractive indexes and between the optical constants along the *b*- and *c*-axis. As a result, the measured in-plane birefringence of GeS$_2$ reaches Δ*n* ~ 0.12. Utilizing the crystallographic data acquired through XRD analysis, we calculated the anisotropic permittivity tensor using first-principle calculations (see Methods and Supplementary Note 2). As shown in Figure 3a, the experimentally measured refractive index and birefringence are in precise agreement with the theoretical estimates.

For additional verification, we performed polarized micro-transmittance measurements of GeS$_2$ flakes exfoliated on Schott glass substrates (see Methods and Supplementary Note 5). Figure 4b shows exemplified transmittance spectra maps for parallel polarizations for 151 nm-thick flake. Then, we calculated the transmittance spectra based on the optical constants (see Supplementary Note 8). Figure 3b,c shows the experimental and calculated transmittance maps in parallel-polarized configuration. We note that calculated spectra perfectly match the experimental data, affirming the reliability of the measured optical constants. Remarkably, the results yield the extinction coefficient *k* close to zero for wavelengths above 362 nm. Hence, our quantitative study extends the list of lossless (bandgap > 3.4 eV) high-refractive-index vdW materials. Meanwhile, GeS$_2$ possesses in-plane and natural vdW out-of-plane anisotropy, as shown in Figure 3a,d. Both in-plane and out-of-plane birefringent properties, in combination with the transparency in the ultraviolet-visible range, distinguish GeS$_2$ among the broad family of vdW materials.

**Near-field imaging of waveguide modes in GeS$_2$ flakes**

To unambiguously verify the extracted dielectric function, we probed propagating modes in GeS$_2$ using scattering-type Scanning Near-field Optical Microscopy (s-SNOM) in the reflection mode (see Figure 4a). A vertically oscillating metalized atomic force microscopy (AFM) tip was illuminated with *p*-polarized light using a parabolic mirror while scanning the GeS$_2$ flake. Acting as an antenna, the mirror concentrated the incident field at the very tip, forming a nano-sized focal spot for local study of material properties and excitation of polaritons. The radiation scattered by the tip and edges of the flake was recorded simultaneously with the topography, enabling near-field imaging with nanoscale resolution. At the same time, a pseudo-heterodyne interferometric module made it possible to recover both the phase and amplitude information from near-field radiation. We performed s-SNOM measurements at three representative wavelengths (700 nm, 850 nm, 1000 nm) on a freshly exfoliated 215 nm-thick GeS$_2$ flake placed on top of the Schott glass substrate.

Prior to the examination of the acquired results, we determined in-plane axes orientations by polarized Raman spectroscopy as discussed in previous sections. Figure 4b-d shows topography and near-field amplitude Amp(*E*) and phase Arg(*E*) signal maps recorded at the wavelength of 1000 nm. In the *s*-SNOM experiment, we observed multiple bright fringes parallel to all the flake edges. The observed fringes result from the interference between two signals: (1) the wave excited at the probe tip that propagates to the sample edge and scatters, and (2) the background signal, which is the result of the reflection of the evanescent field from the sample surface at the probe-sample interface. This interference pattern creates alternating regions of constructive and destructive interference, manifesting as the visible fringe pattern in the experimental observations. In the frequency domain, this background signal corresponds to the zero-frequency peak in the Fast Fourier Transform (FFT) of the fringe pattern. Importantly, the strong dependence of the fringes period on the direction is not only due to the anisotropy of the material but also due to the spatial frequency shift effect arising due to the geometry of the experiment.

To derive effective indices of propagating modes in GeS$_2$ flake, we analyzed the Fourier transform (FT) of complex near-field amplitude built upon both Amp(*E*) and Arg(*E*) line profiles extracted from experimental maps (see Figure 4e,f). In our analysis, we utilized the established fact that modes propagating in different orientations within the flake exhibit spatial frequency shifts for s-SNOM measurements in reflection mode[57,58]. These shifts are intricately linked to the orientation of the flake's edge that scatters the propagating mode and described by the following formula:

$$n_{\text{eff}} = n_{\text{obs}} + \cos(\alpha)\cos(\beta) \qquad (1)$$

where $n_{\text{eff}}$ is the actual effective index of the mode, $n_{\text{obs}}$ is the observable effective mode index, α is the angle between the wavevector *k* of the incident light and its projection of the sample surface $k_{\parallel}$, and *β* is the angle between $k_{\parallel}$ and the direction of propagation of the mode in the sample. Within the recorded near-field maps, we successfully identified both transverse electric (TE) and transverse magnetic (TM) fundamental modes within the GeS$_2$ flake, alongside the presence of air modes. To validate this observation, we performed transfer-matrix calculations[59] based on the previously determined dielectric function of the material. Figure 4g,h indicates calculated energy ($E_{\text{ph}} = hc/\lambda_{\text{vac}}$ with $\lambda_{vac}$ being the wavelength of light in vacuum) versus in-plane momentum ($q_{a,b} = 1/\lambda_{a,b}$ with $\lambda_{a,b}$ being the wavelength of a guided mode) dispersion relation for waveguide modes in a planar waveguide of thickness 215 nm placed on top of the glass. The experimentally determined effective mode indices for both in-plane directions are shown in dispersion maps as green dots. The agreement between experimental and calculated values serves as a robust confirmation of the constants derived in preceding chapters.

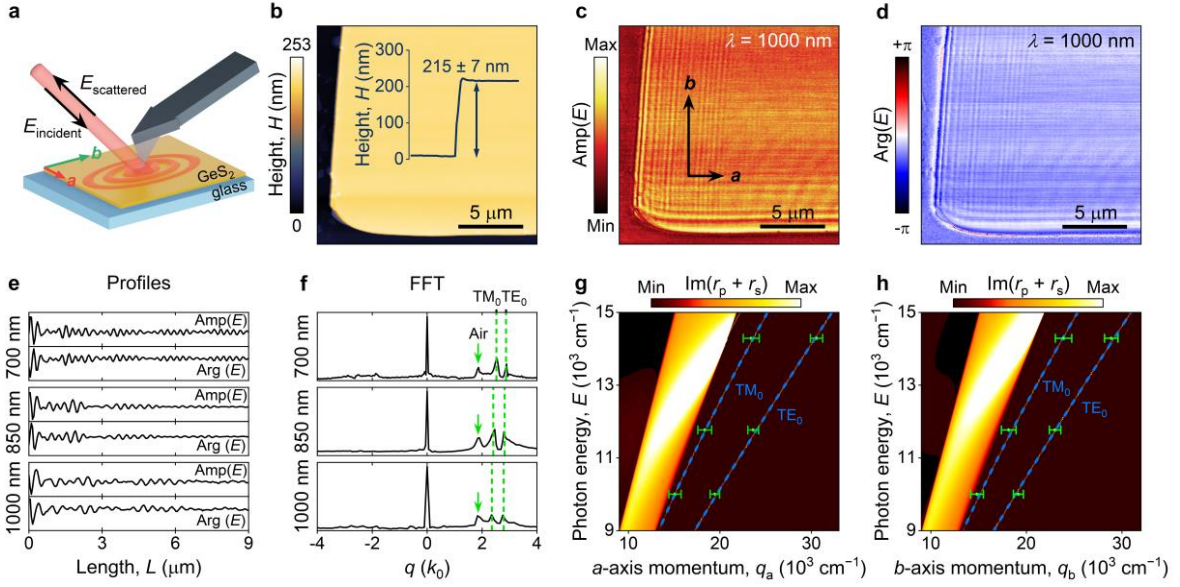

**Figure 4. Confirmation of optical anisotropy of van der Waals GeS$_2$ by near-field measurements.** (**a**) Schematics of the *s*-SNOM experimental configuration for GeS$_2$ sample on top of Schott glass substrate. (**b**) AFM image of the studied GeS$_2$ flake's corner. The inset shows the cross-section height plot. (**c**, **d**) Near-field images: amplitude Amp(*E*) and phase Arg(*E*) of the electric field at the incident wavelength of $\lambda$ = 1000 nm (the other wavelength images are displayed in Supplementary Note 7). (**e**) The line profiles are taken from (c) and (d). (**f**) Fourier transform amplitude of the complex near-field signal presented in (e). Transfer-matrix calculations for propagating modes GeS$_2$ flake along the *a* (**g**) and *b* (**h**) crystallographic directions. Dashed blue lines denote TM and TE fundamental modes. The green dots indicate experimental data points obtained from s-SNOM imaging. In calculations, we used the permittivity extracted from the ellipsometric measurements as an input.

## Discussion

In summary, we present a detailed analysis of the structural and optical properties of exfoliated GeS$_2$ flakes. Using a phonon-based approach, we calculated the vibrational response of the whole structure involving the simultaneous vibration of several structural units and their interactions. Combining polarized non-resonant Raman spectroscopy with *ab initio* calculations we revealed the vibrational origin of 13 lines and identified 5 well-resolved lines whose polar plot maxima and minima are aligned with crystallographic axes. Furthermore, we found that GeS$_2$ demonstrates the highest refractive index among transparent van der Waals materials in the visible and near-ultraviolet regions. It exhibits a remarkably large in-plane and out-of-plane birefringence up to 0.12 and 0.63, respectively. As a result, GeS$_2$ is a rare example of a highly refractive material with a transparency band spanning over the whole visible spectral range. Such materials are very promising for use in integrated photonic nanostructures for light manipulation, *e.g.*, waveguide networks and metasurfaces, and for macroscopic-scale structures, such as waveguide combiners for AR/VR headsets, for example. The former benefits from improved field confinement due to the high refractive index and optical anisotropy[3,25]. In the case of waveguide combiners, the wave nature of light limits the optical resolution to $R < \pi n d \cos(\alpha)/(90\lambda)$ pixels per degree, where *d* is the thickness of the waveguide, *n* is the refractive index of waveguide material, $\lambda$ is free space wavelength of light, and $\alpha$ is the angle between the waveguide surface and the ray. Hence, waveguides based on high-*n*-materials can be made thinner, lighter, and easier to wear. Also, higher *n* extends the range of possible propagation angles $\alpha$, which leads to a greater field of view[60]. More advantages, such as higher diffraction efficiency and suppressed rainbow effect, are gained from the use of highly refractive material in waveguide in- and out-couplers[60,61]. As a result, our findings of GeS$_2$ promising optical properties expand the engineering capabilities for van der Waals materials-based photonics.

**Methods**

**Sample preparation.** Bulk synthetic GeS$_2$ crystals were purchased from SixCarbon Technology (Shenzhen, China) and exfoliated on top of Si, Si/SiO$_2$, and Schott glass substrates at room temperature using commercial scotch tapes from Nitto Denko Corporation (Osaka, Japan). Prior to exfoliation, the corresponding substrates were subsequently cleaned in acetone, isopropanol alcohol, and deionized water, and then subjected to air plasma to remove the ambient adsorbates. To perform TEM analysis, GeS$_2$ flakes were transferred onto TEM membranes (EMS non-porous Si TEM grid) using a polymer-assisted dry-transfer technique performed with polydimethylsiloxane (PDMS) films.

**Raman measurements.** Raman spectra were acquired with a Horiba LabRAM HR Evolution (HORIBA Ltd., Kyoto, Japan) confocal Raman microscope. 100× (N.A. = 0.90) microscope objective and 1800 lines/mm diffraction grating were used. The studies were carried out at 532 nm and 632.8 nm excitation wavelengths. For 532 nm, excitation power density was 60 kW/cm$^2$, and integration time was 120 s for each point. For 632.8 nm, excitation power density was 11 kW/cm$^2$, integration time was 200 s for each point. In both cases, the spot area was 1.8 µm$^2$. The processing of the spectra was carried out as follows: first, the spectra background was subtracted. Afterward, background-subtracted spectra were fitted by Lorentzian lines. Finally, the results of the fitting were used to fit the data and construct the polar graphs. MagicPlot was used to fit the spectra, while OriginPro was used for background subtraction, polar plot fitting, and figure construction. Throughout the studies, angle-resolved Raman spectra were excited by various laser powers and acquired from multiple flakes to ensure the consistency and repeatability of obtained results.

**Spectroscopic ellipsometry.** To analyze the anisotropic optical response of GeS$_2$, we used a commercial imaging spectroscopic ellipsometer Accurion nanofilm_ep4 in the nulling mode. Ellipsometry spectra were recorded for three samples with various thicknesses of *t* = 66 nm, 87 nm, and 128 nm in the spectral range from ultraviolet (250 nm) to near-infrared (1650 nm). For ellipsometry analysis, we followed an algorithm similar to the procedure described in [2] of our recent work. However, for a biaxial material, such as GeS$_2$, the key difference was the alignment of the sample in such a way that one of its in-plane optical axes was perpendicular to the plane of incidence. This was crucial for the separation of s- and p-polarized light, and therefore, Mueller-matrix measurement and analysis were not necessary for measuring the in-plane anisotropy of GeS$_2$.

**Micro-transmittance.** Spectroscopic transmittance was measured using a Zeiss Axio Lab.A1 optical microscope equipped with a halogen light source. Coupling to an Ocean Optics QE65000 grating spectrometer was achieved through a Thorlabs M92L02 optical fiber with a 200 µm core diameter. Transmitted light was collected from a spot smaller than 15 µm, utilizing the "N-Achroplan" 50× Pol

M27 objective with a numerical aperture of 0.8. For a more detailed description of this micro-transmittance setup see Reference[62].

**Atomic force microscopy.** The thickness of GeS$_2$ flakes was accurately characterized by an atomic force microscope (NT-MDT Ntegra II) operated in contact mode at ambient conditions. AFM measurements were obtained employing silicon tips (ETALON, HA_NC ScanSens) with a spring constant of 3.5 N/m and a resonant frequency of 140 kHz. Image processing and quantitative analysis were performed in Gwyddion software.

**Scanning Near-Field Optical Microscopy.** Near-field measurements were performed using a scanning near-field optical microscope ("NeaSNOM" from company Neaspec GmbH). Ti:Sapphire continuous wave tuning laser (TiC, AVESTA Lasers, and Optical Systems) with fiber coupling output was used, working at a wavelength range of $\lambda$ = 700 – 1000 nm. We used Pt-coated silicon tip oscillating at a resonance frequency of $\Omega \approx$ 280 kHz with an amplitude of ~80 nm (ARROW-NCPt-50). The microscope was used in the reflection mode, meaning the same parabolic mirror was used to excite and collect near-field signals. To enhance the quality of the near-field image, we reduced the optical background by demodulating the received signal at a high-order harmonic frequency $n\Omega$ (where $n$ can be 2, 3, or 4), and using an interferometric pseudoheterodyne approach with a reference beam modulated by an oscillating mirror. For this particular study, the signal demodulated at the third harmonic (3$\Omega$) proved adequate for analysis free from background interference.

**X-ray diffraction analysis.** X-ray diffraction analysis of GeS$_2$ single crystal was performed on a Bruker D8 QUEST diffractometer with a Photon III CMOS area detector using Mo K$\alpha$ radiation ($\lambda$ = 0.71073 Å) focused by a multilayer Montel mirror. The full data set was collected at 300 K within two $\omega$-scans applying a sample-to-detector distance of 60 mm. Raw data were indexed with cell_now and integrated using SAINT from the APEX4 package[63]. Absorption correction was performed using a numerical method based on crystal shape as implemented in SADABS[64]. The crystal structure was solved by direct methods using SHELXS and refined anisotropically with the full-matrix F2 least-squares technique by SHELXL[65,66]. CSD reference number 2363955 contains supplementary crystallographic data for this paper. These data can be obtained free of charge from the Cambridge Crystallographic Data Centre *via* https://www.ccdc.cam.ac.uk/data_request/cif.

**First-principle calculations.** *Ab initio* calculations were performed using the Vienna *ab initio* simulation package (VASP) code[67]. The core electrons were described with projector augmented wave (PAW) pseudopotentials treating the Ge *3d*, *4s,* and *4p* and the S *3s* and *3p* electrons as valence electrons. A kinetic energy cutoff for the plane-wave basis was set to 450 eV. For all of the computations, the lattice parameters, as well as atomic positions, were taken from our XRD measurements. Raman spectra and polar diagrams were computed based on the Phonopy-Spectroscopy package with our modifications. Optical properties of GeS$_2$ were calculated within GW approximation on top of wavefunctions calculated using PBE functional using the VASP code. For this, we used $\Gamma$-centered *k*-points mesh sampling the Brillouin zone with a resolution of 2$\pi \cdot$0.05 Å$^{-1}$.

## Author Contributions

A.S.S., G.A.E., I.A.Z., D.V.G., and K.V.K. contributed equally to this work. G.A.E., G.I.T., A.V.A., V.S.V., and K.S.N. suggested and directed the project. A.S.S., G.A.E., I.A.Z., D.V.G., M.K.T., M.S.M., A.N.T., M.R.P., A.V.S., D.I.Y., D.M.T., D.A.G., and S.M.N. performed the measurements and analyzed the data. K.V.K., I.K., K.V.V., I.M.F., and A.A.V. provided theoretical support. A.S.S., G.A.E., I.A.Z., K.V.K., and A.A.V. wrote the original manuscript. All authors reviewed and edited the paper. All authors contributed to the discussions and commented on the paper.

## Competing Interests

The authors declare no competing financial interest.